# Carbogels for sustainable and scalable thermoelectric applications

Shoeb ATHAR[1,2], Jérémy GUAZZAGALOPPA[1,2], Fabrice BOYRIE[1], Cédric HUILLET[2] and Philippe JUND[1,*]

1. ICGM, Univ Montpellier, CNRS, ENSCM, 34293 Montpellier, France
2. HUTCHINSON SA - Rue Gustave Nourry, 45120 Châlette-sur-Loing, France
* Corresponding author. E-mail address: philippe.jund@umontpellier.fr (Philippe JUND)

**ABSTRACT**

Thermoelectric generators (TEGs) based on commercially used thermal super-insulating materials can facilitate sustainable and large-scale ambient waste heat recovery while bequeathing an added economic and environmental value to thermal insulations in industry. This requires the optimization of the thermoelectric (TE) properties through electrical functionalization of such materials. Moreover, the associated engineering challenges of assembling TEG modules must be overcome. Herein, we propose using super-insulating Resorcinol-formaldehyde (RF) carbogels for scalable and sustainable TE applications through their electrical functionalization. Using a combination of a pyrolysis process and carbon fibers insertion, we achieved an increment by 12 orders of magnitude in electrical conductivity as well as *ZT* whilst retaining their intrinsic ultralow thermal conductivity (<50 mW/mK). A TE module in the form of a thermoelectric vacuum insulation panel (TVIP), was then fabricated using only a p-type material, to demonstrate a proof-of-concept self-powered WiFi-based vacuum-failure detection application in confined spaces in automobiles or aeronautics. Finally, by extrapolating the optimized output power and with a CAD-assisted assembly of a large TEG module (1000 cm$^2$), the potential of scalable low-grade waste heat recovery is discussed.

## 1. Main

Climate change and the increased global power demand is further exacerbated by the unabated depletion of fossil fuel resources, thus making this world an entity craving for energy and simultaneously struggling for its existence due to the consequences of fossil fuel energy consumption [1]. Since more than 2/3rd of the used fossil energy is lost as waste heat, development of thermoelectric generators (TEGs), typically consisting of p-type and n-type thermoelectric (TE) materials which produce electricity directly from heat--employing the Seebeck effect - is highly essential [2]. The performance of a TE material is rated by the dimensionless figure of merit **ZT** given by [3] **ZT = ($S^2\sigma$) ×T/ $\kappa$** (where **S** is the Seebeck coefficient (V.K$^{-1}$), **T** is the working temperature (K), $\sigma$ is the electrical conductivity (S.m$^{-1}$), and $\kappa$ is the thermal conductivity (W.m$^{-1}$.K$^{-1}$). A good TE material has both high electrical conductivity and Seebeck coefficient – coupled together as $S^2\sigma$ in the TE power factor (PF) – while $\kappa$ should be minimal.

At present, waste heat at high temperatures (> 673 K) or medium temperatures (373–673 K) has been considered valuable for TE electricity conversion or in other industrial processes [4]. Large quantities of low temperature heat (< 350 K), which constitutes more than 70% of available industrial waste heat [5], remains unused for TE conversion due to one or more of the following reasons: scarcity of efficient room-temperature TE materials [6]; presence of scarce, toxic, and expensive elements (e.g. $Bi_2Te_3$, PbTe, etc) [7]. While some materials may filter through the checklist thanks to the tremendous advances in the TE materials research, large scale commercialization and practical applications of TEGs have proceeded very slowly due to a myriad of roadblocks [8]. Foremost, there are engineering challenges such as weight reduction for mobile TE application; thermal and electrical interface degradation due to chemical instability (e.g. $Cu_2Se$) [6]; microcrack formation and limited cutting geometries/flexibility due to brittleness (e.g. chalcogenides); thermal stresses due to thermal expansion mismatch between n-type and p-type legs for many efficient TE materials (e.g. $Zn_4Sb_3$) that can only be synthesized as one of the types [9] etc. Then there are commercial challenges arising out of scaling and automating the production of TE materials as well as manufacturing and assembly of modules. It is for these reasons that sustainable and scalable use of TEGs has been hindered by a cost-efficiency tradeoff [10].

While, ultimately, the ongoing research on TE materials/TEGs may find a perfect solution, climate emergency warrants urgent action exploiting a broad portfolio of affordable and achievable strategies [11]. One such strategy involves electrically functionalizing commercially used thermal super-insulating materials by doping methods aimed at increasing the concentration of charge carriers [12]. Such materials are nontoxic, low density, sustainable, and low-cost with scalable production [13, 14]. With an intrinsically and extremely low thermal conductivity, a small doping-induced optimization of electrical conductivity and Seebeck coefficient, or power factor per se, can significantly enhance their TE performance by several orders of magnitude [15]. Since such materials are already being commercially used at large scale for insulation, in industries, automobiles, etc., such an electrical functionalization can bestow them with a complementary function which may be conveniently exploited [12]. Consequently, any amount of waste heat harnessed will prove to be of additional commercial and environmental value beside the primary super-insulating application of these materials.

Resorcinol-formaldehyde gels (aerogels/xerogels) are intensively studied for thermal insulation application thanks to their ultralow thermal conductivity. Their manufacturing process is less expensive than the usual materials synthesis methods and opens the door to previously unattainable thermoelectric module architectures-with modular shapes (surfaces/thicknesses) exhibiting flexibility properties [16] - given their electrical conductivity can be improved.

In this work, we first considered improving the *ZT* of this material by converting it into carbon xerogel (carbogel) through a pyrolysis process, and with subsequent insertion of conductive fibers. Thereafter, using only a *p*-type TE material, a flexible TE module in the form of a thermoelectric vacuum insulation panel (TVIP), was fabricated to demonstrate a proof-of-concept wireless vacuum-failure detection application for the development of self-powered sensors in confined spaces in automobiles or aeronautics. Finally, through extrapolation of optimized output power and CAD-assisted fabrication of large TE modules (1000 $cm^2$), the prospects of large-scale application for ambient waste heat recovery are discussed.

## 2. Results

**2.1 Study of the thermoelectric properties of carbon RF carbogel**

Table 1 shows the TE properties of the pristine and pyrolyzed RF carbogel whereas the influence of pyrolysis on their structural properties is discussed in section S2.1 of the supplementary information (SI). Clearly, the non-pyrolyzed material is an electrically insulating material and thus has the lowest figure of merit *ZT* (300 K) ≈ $10^{-16}$. Pyrolysis renders the material semi-conducting by increasing its electrical conductivity by 11 orders of magnitude (from $\sigma = 10^{-10}$ $S.m^{-1}$ to $\sigma = 10$ $S.m^{-1}$) which explains the dazzling improvement in its figure of merit, with *ZT* (300 K) ≈ $10^{-5}$ . Increasing the pyrolysis temperature can still result in higher electrical conductivity. However, it is not recommended to go above 850°C for the conductive fibers we would subsequently use (Basalt and PANOX) as it causes shrinkage as well as large deformations of the fiber, which generally leads to cracking of the material after pyrolysis. We also note that pyrolysis only slightly deteriorates the thermal insulation properties of our material, since the thermal conductivity remains very low, with $\lambda < 32$ mW.m$^{-1}$ K$^{-1}$ thanks to the high porosity of the material. It is the same for the density, which in fact increases slightly with the pyrolysis process but also remains low, $\rho < 87$ $kg.m^{-3}$.

| Materials | Properties | | | | |
|---|---|---|---|---|---|
| | λ (mW. m$^{-1}$K$^{-1}$) | σ (S.m$^{-1}$) | α (μV.K$^{-1}$) | ZT (300 K) | ρ (kg.m$^{-3}$) |
| RF Xerogel | 25 | 1E-10 | 15 | 2.7E-16 | 70 |
| RF P$_{850°C}$ | 30 | 10 | 17 | 2.9E-5 | 85 |

Table 1: Thermoelectric properties of non-pyrolyzed RF xerogel and pyrolyzed RF Carbogel at 850°C (RF P$_{850°C}$)

The pyrolysis process makes it possible to improve the electronic properties of the RF xerogel due to the monolithic structure of the resulting carbogel composed of carbon particles linked by covalent bonds. This means that the electrical conductivity is directly related to the connection between particles. In addition, high temperature treatment further increases the electrical conductivity. This increase can be attributed to the appearance of new layers of graphene as well as lower contact resistances between particles [17]. Due to thermal activation, electrical conductivity is governed by two mechanisms, namely: the movement of charge carriers along the network of the material by inter-aggregate contact, or a transfer of charge carriers from one electrically conducting segment of the network to another separated by an amorphous zone. In these cases, an electron hopping mechanism or a tunneling mechanism allows electrons to cross potential barriers [18, 19]. It should be noted that conduction by tunneling applies when particles are separated by a distance of only a few nanometers. The conduction electrons can thus pass the energy barrier allowing them to pass from one conductive particle to another in the polymer matrix. In addition, an increase in temperature makes it possible to provide additional energy to the electrons, thus promoting conduction mechanisms by electron hopping and tunneling [20]. These different mechanisms ensure continuous diffusion of charge carriers, even in the case where there is a break in the links at the level of the network chains. This, therefore, contributes to a high electrical conductivity by maintaining the so called 'percolation' of the system. The theory of percolation explains the percolation phenomenon through several statistical models [21, 22] by dividing it into three zones (shown in figure 1) as follows:

- Zone 1: This zone corresponds to a low charge rate $\varphi < \varphi_1$. The conductive charges are isolated in the gel matrix and do not interact with each other. In this zone, the electrical conductivity depends only on the electrical conductivity of the gel alone.

- Zone 2: This zone corresponds to the percolation zone. Due to the increase in the conductive charge content, interactions between particles are more frequent, which allows the appearance of the tunnel conduction mechanism. Thus, several conduction paths are created, explaining the sudden increase in the electrical conductivity of the material. This zone is called the percolation threshold, it is generally associated with a critical conductive charge content $\varphi_c$, such that $\varphi_1 < \varphi_c < \varphi_2$.

- Zone 3: This zone is located beyond the percolation zone. For a charge content greater than $\varphi_2$, the electrical conduction plateau is reached. Thus, a further increase in the conductive charge content has only a very small influence on the electrical conductivity.

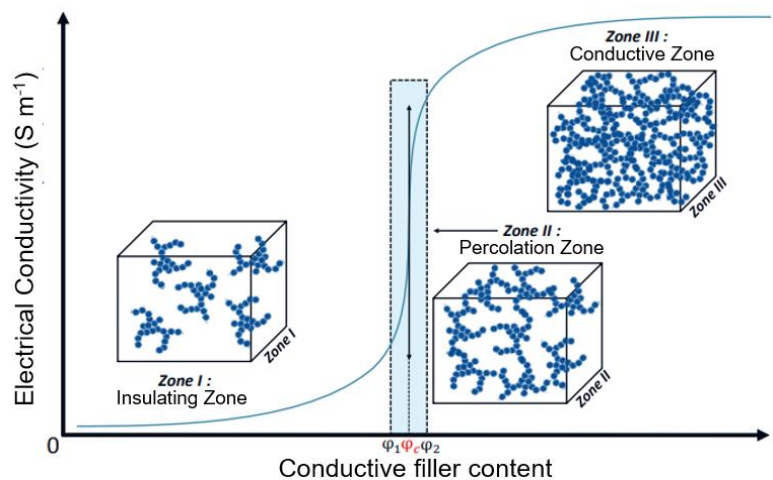

Figure 1: Diagram illustrating the theory of electrical percolation as a function of the content of conductive charges

Though in this case, even if we managed to obtain a semi-conducting material and a good improvement in electrical conductivity, we note that the figure of merit remains low, with *ZT* $(300\ K) \approx 10^{-5}$. This is partly due to the low Seebeck coefficients with $\alpha < 17\ \mu V.K^{-1}$. We can, therefore, conclude that the thermoelectric properties of the material still need to be improved. To achieve this, we subsequently used the insertion of charges (carbon fibers). The goal was then to identify a type of conductive charges that would allow us to optimize the thermoelectric properties of our material, while maintaining its thermal insulation property and low density.

## 2.2 Doping the RF carbogel with different conductive fibers

The choice of conductive charges for this study was based on their form factor defined as the ratio of the maximum ($D_{max}$) and minimum ($D_{min}$) Féret diameters [23]. This geometric parameter influences the quantity of charges necessary to reach a percolation threshold and thus create a conduction path throughout the volume of the material. A charge is said to have a low form factor if its geometry leads to $D_{min} \approx D_{max}$ or a high form factor if its width is much smaller than the length. We chose Basalt and oxidized polyacrylonitrile (PANOX) nanofibers due to their high form factor which would make it possible to form a conduction path at very low doping rates. Due to high electronic mobility, their electrical conductivity is generally of the same order of magnitude as that of graphite, around $\sigma \approx 10^4\ S.m^{-1}$ [24]. Further, they are arranged quite easily in the three-dimensional network of organic gels and reinforce the mechanical properties of the RF carbogel [25, 26].

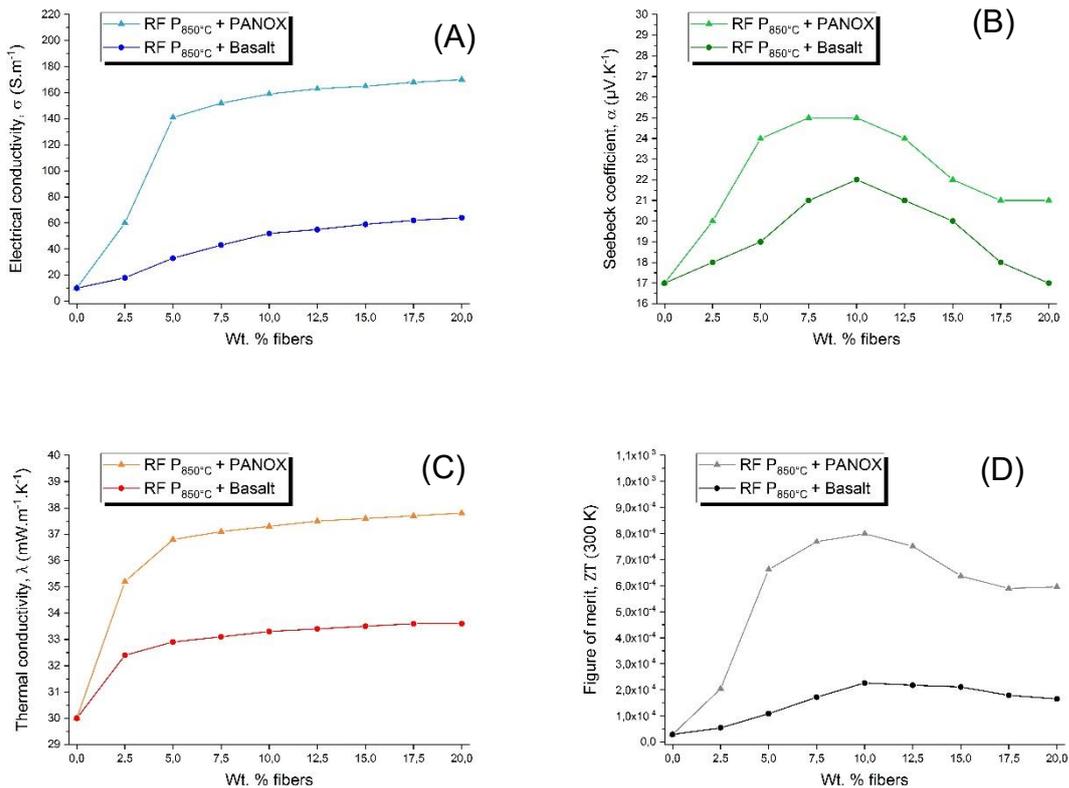

Figure 2: Study of the thermoelectric properties of RF carbogels pyrolyzed at 850°C as a function of the content of basalt fibers and PANOX fibers, with (A) electrical conductivity, (B) Seebeck coefficient, (C) thermal conductivity and (D) *ZT* (Figure of merit)

Figure 2 highlights the improvement in thermoelectric properties as a function of basalt fibers and PANOX fibers content. According to these results, PANOX fiber results in a thermoelectric performance approximately four times greater than that obtained with basalt fiber. This is mainly explained by the fact that PANOX fiber has a much higher electrical conductivity than basalt fiber, with $\sigma \approx 10^3$ Sm$^{-1}$ for basalt and $\sigma \approx 10^5$ Sm$^{-1}$ for PANOX. The Seebeck coefficient is slightly higher with PANOX fiber ($\alpha \approx 25$ µV.K$^{-1}$ with 10% fiber) than with basalt fiber ($\alpha \approx 22$ µV.K$^{-1}$ with 10% fiber). However, we note that the thermal conductivity of the final material is higher in the case of the PANOX fiber ($\lambda \approx 37$ mW.m$^{-1}$.K$^{-1}$ with 10% fiber) than with the basalt fiber ($\lambda \approx 33$ mW.m$^{-1}$.K$^{-1}$ with 10% fiber). Nevertheless, the values of the thermal conductivity and the Seebeck coefficient of these two fibers remain similar, compared to the difference in electrical conductivity, thus explaining such a difference in the figure of merit. Noticeably, beyond a doping rate of 10%, while the electrical and thermal conductivity continue to increase - although marginally for the former; the Seebeck coefficient decreases. Since, the Seebeck coefficient tends to decrease with the charge carrier concentration [27], it is difficult to explain the reason for the simultaneous increase in electrical conductivity and Seebeck coefficient. However, this phenomenon, already has some precedence in previous research works [28-32]. In our case, the reason for the increase of the Seebeck coefficient with the doping rate can be explained by the appearance of a potential barrier. It can be attributed to the possibility that the work function (minimum energy required to remove an electron from the surface of a metal) of RF xerogel during pyrolysis is smaller than that of pyrolyzed carbon [33, 34], thus forming a potential barrier. Hence, charge carriers cannot easily move between the graphene sheets and the pyrolyzed RF matrix, due to the presence of this energy barrier. Therefore, only charge carriers with higher energy than the potential barrier can move, while those with lower energy are scattered, which leads to the average decrease in the charge carrier density and therefore the increase in the Seebeck coefficient [35-41]. It should also be noted that the simultaneous increase in the Seebeck coefficient and the electrical conductivity is characteristic of the presence of the electron tunneling or hopping mode as already explained in section 2.1. and attributed to the presence of a percolation threshold.

On the other hand, for higher doping rates (> 10%), the percolation threshold is exceeded, in addition, the potential barrier is significantly weakened due to the excessively high concentration of charge carriers, which explains both the stagnation in electrical conductivity and the reduction in the Seebeck coefficient [28, 30]. Thus, in the continuation of this work, we didn't persist with higher doping rates - hypothetically making it possible to reach the percolation threshold- since the material, otherwise, would be neither a good thermal insulator nor a good thermoelectric material, but only a good electrical conductor. The densities of the RF P850°C carbogels containing 10% basalt fibers and 10% PANOX fibers remain low at $\rho_{RF+basalt}$ = 140 kg.m$^{-3}$ and $\rho_{RF+PANOX}$ = 130 kg.m$^{-3}$, respectively. Precisely, both fiber materials have a maximum figure of merit with a fiber mass content of approximately 10%: $ZT$ (300 K) = 8 × 10$^{-4}$ for the RF carbogel P850°C with PANOX fibers and $ZT$ (300 K) = 2.3 × 10$^{-4}$ for the RF P850°C carbogel with basalt fibers. While this value is still few orders of magnitude lower than the $ZT$ of state-of-the art TE materials, as we will show in the subsequent sections, it is sufficient to provide the few µW/mW necessary for wireless sensor applications [42] yet with a significant advantage in weight.

The SEM images of the gel matrix illustrate the arrangement around the basalt fibers and the PANOX fibers (figure 3). From these images, we see that the RF matrix tends to cover the fibers. Thus, these fibers also play the role of mechanical reinforcement which allows homogeneous migration of the RF gel all around them. In addition, at iso-thickness, PANOX fiber ($\rho$ = 450 g.m$^{-2}$) has a lower surface density than basalt fiber ($\rho$ = 480 g.m$^{-2}$) which promotes capillary permeability by reducing surface tensions between the fibers and the RF gel [43]. This also explains the better mechanical strength noted with the PANOX fiber, at the end of the pyrolysis process.

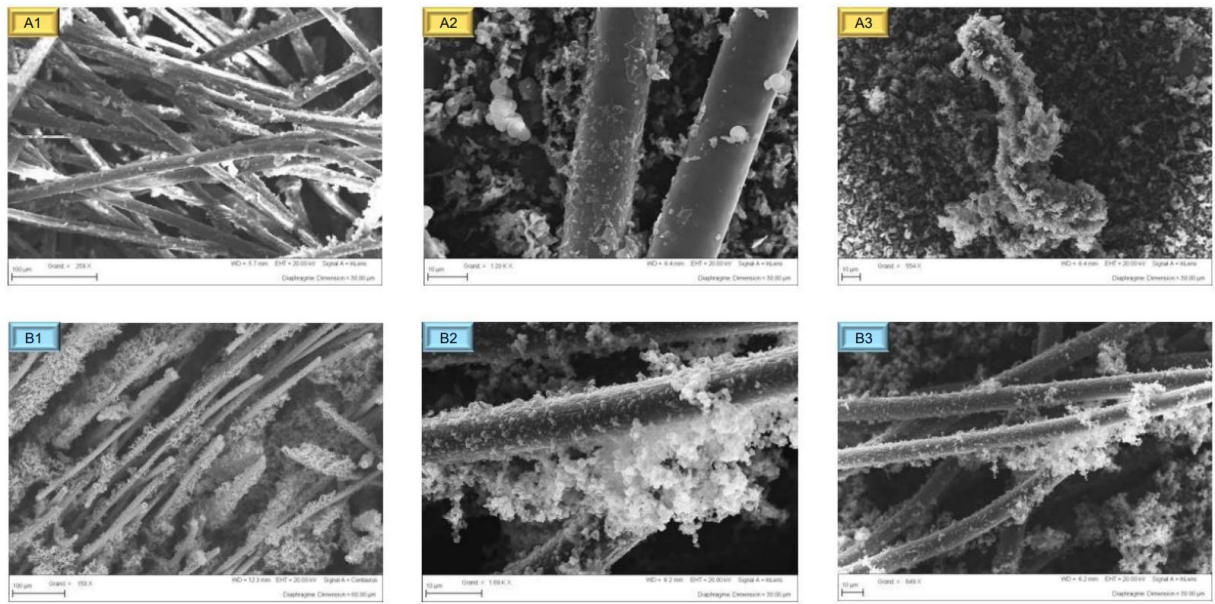

Figure 3: SEM images of RF carbogels pyrolyzed at 850°C with a content of 10% in basalt fibers (A1, A2, and A3) and with a content of 10% in PANOX fiber (B1, B2, and B3)

## 2.3 Thermoelectric vacuum failure detection system

The first application envisaged for our TE module, is a failure detection system to detect the deterioration of the thermal insulation of a thermoelectric vacuum insulation panel (TVIP) due to vacuum loss. As an example, we select the thermal insulation of a battery in a hybrid vehicle which has a cooling system maintaining the face of the TVIP against the battery at 10°C. The hot-side temperature of TVIP can be safely assumed to be 50°C due to environmental conditions (vehicle exposed to the sun) or from engine heating in the case of a hybrid vehicle. The TVIP has the advantage of limiting the intake of heat coming from the outside and arriving in the cooling circuit dedicated to the battery. Current pressure, force, or temperature based vacuum sensors require a manual scan [44-46] to recover the data and result in additional mass and costs than our system. Our system, thus, makes it possible, among other things, to limit the oversizing of the cooling circuit, which manifests itself in a saving in space and weight, thus extending the autonomy of the battery. It is based on the principle of a decrease in electrical output of a faulty system upon vacuum loss in TVIP (e.g. due to piercing) as a consequence of a decrease in the thermal conductivity and loss of contact surface from the partial separation of the copper plates against the RF carbogel material.

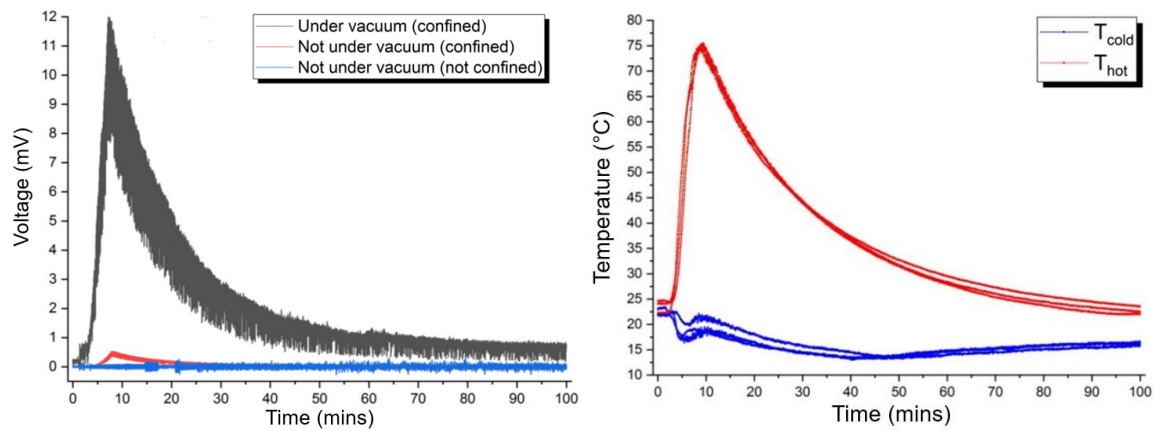

Figure 4: (A) Loss of vacuum test - Measurement of the output voltage as a function of time during a heating cycle, according to three scenarios: a compliant material (TVIP), a failing material (loss of vacuum) subjected to a pressure of 3 kPa and a failing material without mechanical stress (B) Temperature curves associated with the graph representing the evolution of the output voltage as a function of time in figure 4 (A)

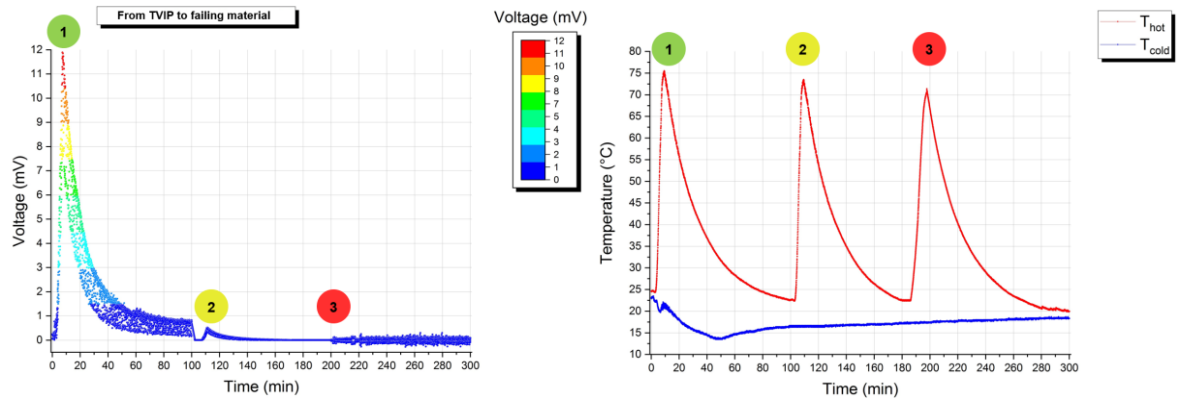

Figure 5: (A) Loss of vacuum test - Measurement of the output voltage as a function of time during three heating cycles, when the TEG is under vacuum (TVIP) and subjected to mechanical stress (cycle 1), then when the TVIP is pierced and subjected to this same mechanical stress (cycle 2) and finally when the TVIP is pierced and without mechanical stress (cycle 3) (B) Temperature curves associated with the graph representing the evolution of the output voltage as a function of time

This paragraph illustrates a case study with the example of the insulation of a battery for a hybrid vehicle with the PANOX fibered RF carbogel pyrolyzed at 850°C. A thickness of 2 mm was finalized for the material after studying the evolution of the surface power as a function of the thickness of the material for several contact resistance values (see section S2.5 and figure S9 in the supplementary information). The material being partly intended for thermal insulation applications, can be assumed to be placed in confined spaces - most often between two walls - exerting a mechanical stress on the material. Thus, to best represent reality, we also tested the utility of our TVIP to detect vacuum failure for a material subjected to mechanical stress. The results of this study are shown in figure 4. It is evident that unlike the case study illustrating the example of the insulation of a battery of a hybrid vehicle, we observe here a drastic reduction of the voltage delivered by the TEG when it loses vacuum. Indeed, for a difference of maximum temperature observed after 10 minutes of acquisition, the voltage goes from 12 mV in the case of the material under vacuum, to zero voltage in the case of a failing TEG without mechanical stress. We also notice that the phenomenon is less pronounced when the material loses vacuum but remains subject to a mechanical stress of a few kPa. These observations can be explained by the fact that when the material loses the vacuum, the plastic envelope will tend to swell slightly which will cause the copper plates to come off; whereas, for TVIP under mechanical stress, the copper plates are not completely detached and still partially fit the material. Yet, the contact resistance is modified and decreases to around 0.5 mV after 10 minutes of acquisition.

Following this study, we carried out the same test using the same material, with the aim of simulating the appearance of a leak occurring in the TVIP. To do this, we started with the material under vacuum constrained by a pressure of a few kPa. We measured the output voltage during the first heating cycle, then we intentionally pierced the TVIP and measured the output voltage again during a second heating cycle. Finally, we removed the mechanical constraint and measured the output voltage one last time during a third heating cycle. We can first notice that the results of this test, shown in figure 5, are similar to those obtained previously: the TVIP material subjected to mechanical stress (zone 1) delivers a maximum of approximately 12 mV during the first heating cycle. Then, when the TVIP is pierced, the voltage drops to approximately 0.5 mV during the second heating cycle. Finally, when we remove the mechanical constraint, we only recover noise at the output due to the separation of the copper plates. But the most interesting observation is the evolution of the output voltage when going from vacuum material to non-vacuum material. In fact, it is at this precise moment that the failure appears. We then notice that at the time of the transition between the TVIP material and the failing material, the voltage drops suddenly, marking a fairly visible break in the curve. So visible that an algorithm would be able to identify the appearance of such a phenomenon. Thus, instead of measuring both the voltage and temperature differences between a vacuum and non-vacuum material, which can sometimes represent an additional constraint, studying the distribution of measurement points corresponding to the output voltage should suffice. A break in the curve, or a significant change in slope, materializes the appearance of the failure, hence, obviating the need for temperature measurements.

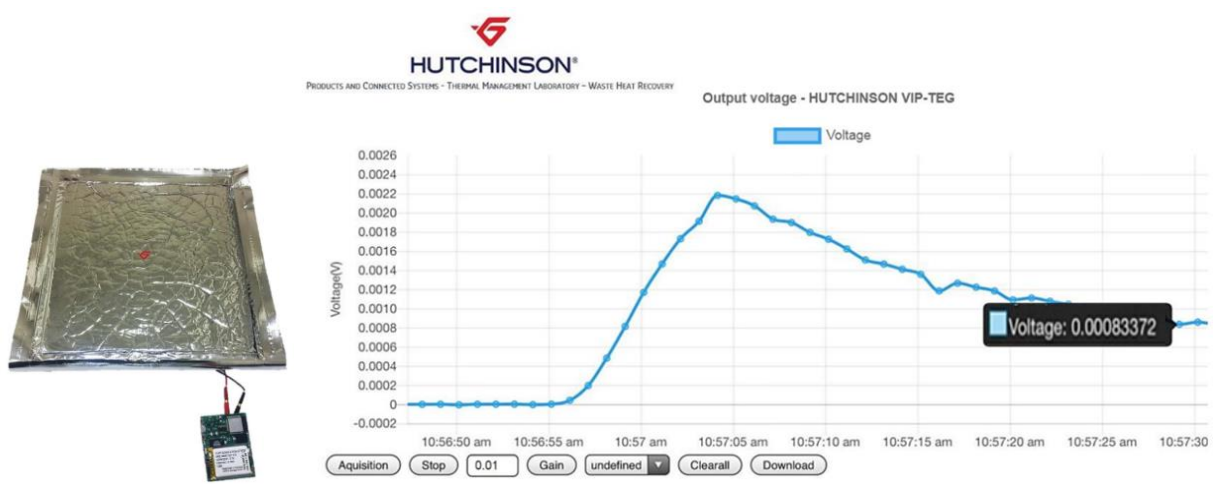

Figure 6: (A) Picture of a thermoelectric functionalized TVIP, connected to the electronic card allowing signal transfer via WiFi (B) HTML interface allowing the visualization and recording of data from the TEG, by WiFi connection to the electronic card

It should also be kept in mind that TVIP, on its own, is only a system for converting thermal energy into electrical energy and is not sufficient to relay output power information indicating a possible failure. A wireless communication system allowing the transmission of data is therefore necessary. Figure 6 illustrates such a system equipped with a means of wireless communication, as part of the thermal insulation of the battery of a hybrid vehicle. For example, we can imagine that the power signal is relayed to the vehicle dashboard, thus alerting the user in the event of a vacuum failure. In order to enable data transfer to warn the user of a possible failure, we used an amplifier system composed of a WiFi antenna, directly connected to the electrical cables at the TVIP output as shown in figure 6.A. An Arduino type electronic card, operating on a rechargeable battery 1 Ah capacity, requiring 200 mA on continuous use and few µA on standby was developed to allow for extended use. This electronic card allows the voltage value delivered by the TVIP to be sent, via WiFi, to devices such as computers, smartphones or tablets. The first step consisted of finding the maximum operating point (MOP) specific to our material, by modifying the gain value, since our card does not have a MOP algorithm unlike some other DC-DC converters. This was followed by measurement repeatability tests consisting of several heating cycles. The results are shown in figures S11 and S12 and discussed in section S1.7 and S1.8 in the supplementary information (SI), respectively.

We then also developed the html page with a secure password connection to allow the acquisition, visualization, and download of the measurement curves of the voltage delivered by the TEG as a function of time as well as to modify the amplifier gain. Thus, after several tests, we chose to display a voltage value every second, corresponding to the average of 4 measurement points in one second (one measurement every 250 ms). Figure 6.B illustrates the interface allowing the visualization and recording of data from the thermoelectrically functionalized TVIP through a device connected via WiFi to the electronic card. In addition, we have identified another means allowing the detection of vacuum failure. This involves comparing the expected electrical powers (calculated from model 1 established in section S1.5) for 1m² of insulation, between a compliant material (TVIP) and a failing material (Figure S13). The results are presented in figure S13 and discussed in section S1.9 of the SI. To sum up, we have demonstrated the feasibility of the TVIP system for failure detection: the output voltage recovered allows us to be informed of a possible loss of vacuum in the thermal insulating material surrounding a battery. It should be noted that compared to a conventional thermocouple, the detection by TVIP is spontaneous upon vacuum loss as it immediately alters the TE properties and causes the copper electrodes to detach thus increasing the contact resistance. A thermocouple will detect the failure by the change in temperature at a given spot and will therefore be subject to thermal inertia, resulting in a slower detection. Furthermore, unlike a thermocouple, our wireless self-powered system can be applied in areas where cables/connections cannot be reliably used.

## 2.4 Large scale application and perspectives

So far, we only used a module with a single p-type TE element. We consider a series association of several pyrolyzed and fiberized RF carbogel type materials in PANOX, with the aim of manufacturing a thermoelectric module with exclusively p-p junctions. The mathematical models established in S1.5 (SI) to calculate the power delivered by the TEG remain valid, provided that the thermal conductance, the electrical resistance and the Seebeck coefficient are multiplied by the number of junctions. Thus, in a second step we defined a "target" material having the thermoelectric properties that we would like to achieve, i.e. a Seebeck coefficient of about 50 $\mu V.K^{-1}$, a thermal conductivity of about 30 $mW.m^{-1}.K^{-1}$, and an electrical conductivity of the order of $10^3$ $S.m^{-1}$. These objectives remain a priori reasonable in view of the current thermoelectric properties of our material. The main difficulty then consists of obtaining an electrical conductivity of the order of $10^3$ $Sm^{-1}$ since it is currently of the order of $10^2$ $S.m^{-1}$. We, however, have already improved this value by more than 12 orders of magnitude, and, hence, we can reasonably expect that it is still possible to improve the electrical conductivity by an order of magnitude more. Furthermore, studies have already revealed electrical conductivity values of thermoelectric organic materials which can reach several hundred $S.m^{-1}$ (or sometimes even 1000 $S.m^{-1}$) and having Seebeck coefficient values greater than 50 $\mu V.K^{-1}$ [28, 30-32] – albeit with too high thermal conductivities for a thermal insulation application. Our unpublished results using reduced-graphene oxide are already promising, though cheaper conductive fillers must be explored for this optimization. Finally, we defined a second target material, like the target material stated previously. The only difference is that the latter has a negative Seebeck coefficient of -50 $\mu V.K^{-1}$. Thus, through these different materials, we calculated the output voltage expected as well as the power density delivered by a module composed of p-p junctions with the current material, then p-p junctions with the target material and finally n-p junctions with the two target materials as a function of the number of junctions and the temperature difference, while neglecting contact resistances. In each case we considered the cold side temperature equal to the ambient temperature as well as a thermoelectric module of 1m² and thickness 2mm. The results of this study are shown in figure 7.

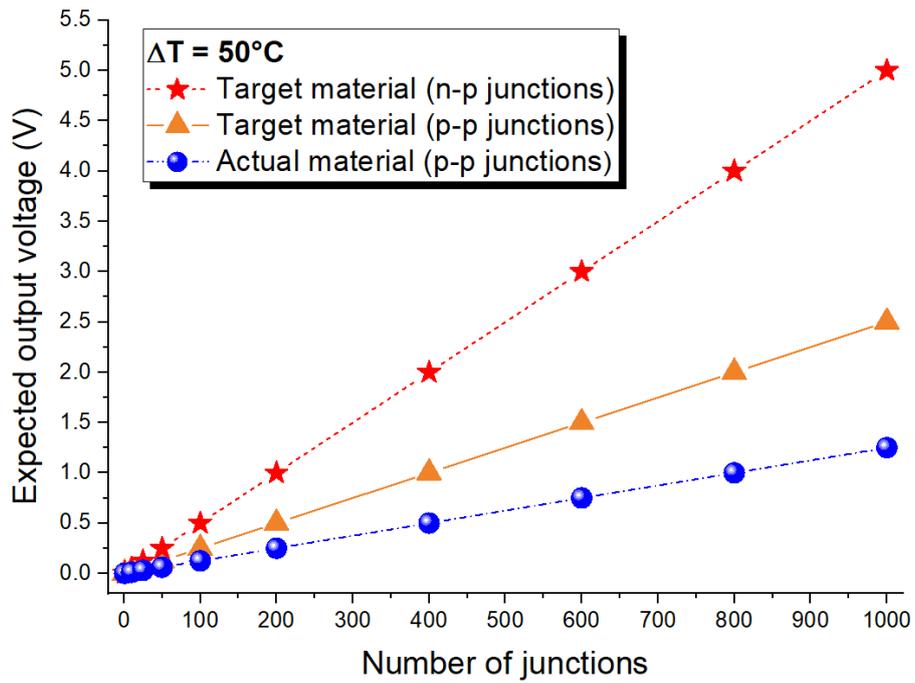

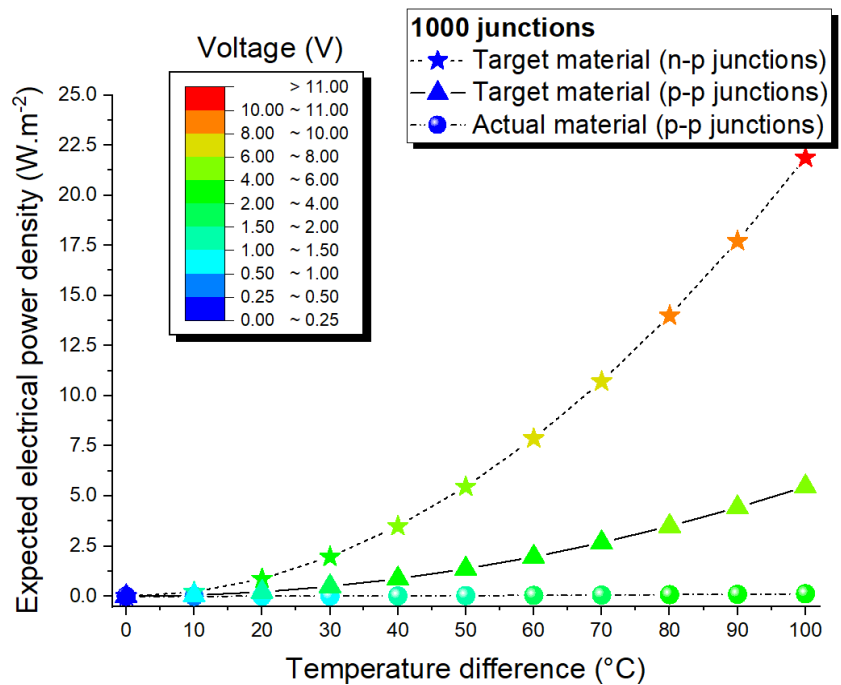

Figure 7: (A) Calculation of the output voltage as a function of the number of junctions with a temperature difference of 50°C, for different thermoelectric modules of 1 m² surface area (B) Calculation of the expected electrical power density and the output voltage as a function of the temperature difference, for different thermoelectric modules composed of 1000 junctions (10 cm² per junction)

Figure 7.A demonstrates a linear variation of the output voltage with the number of junctions. This is explained by the fact that, as in other studies, we consider the contact resistances as perfect. Here we rely on the fact that the vacuum allows to exert mechanical pressure on the material therefore justifying the omittance of these contact resistances. Nevertheless, this first graph perfectly illustrates the interest in moving towards an architecture composed of a large number of junctions: for a difference of 50°C and a 1m² module composed of 1000 junctions (10 cm² per junction), the expected output voltage with the current material reaches 1.25 V, then 2.5 V with the target material (p-p junctions) and finally 5 V with the combination of target materials (n-p junctions). In Figure 7.B, if we now set the number of junctions to 1000 and vary the temperature difference from 0°C to 100°C, then we see that the n-p architecture offers a power density much higher compared to the other two systems: the expected power density for the n-p junction module reaches 22 $Wm^{-2}$ compared to 5.5 $Wm^{-2}$ with the target material with p-p junctions and 125 $mW.m^{-2}$ with the current material, for a temperature difference of 100°C. These values are very interesting for energy storage applications, e.g., aimed at powering different sensors or auxiliaries [42] in order to improve safety or cabin comfort in a car or airplane cabin.

Therefore, for energy recycling applications (recharging a capacitor for example), we have seen through this last theoretical study that an architecture in the form of junctions then becomes necessary. Starting with a computer-aided design (CAD), we finally demonstrated the feasibility of an experimental assembly process for creating such a large module. The CAD representation is illustrated in the figure 8.A. Subsequently, the experimental assembly of such a module was carried out. We assembled 10 grids identical to that presented in figure 8.B, with the aim of reaching 100 junctions. Each pad is separated using an electrical insulating support grid, in order to maintain the final structure during evacuation giving rise to TVIP. This first assembly test on electrical conductivity continuity turned out to be very encouraging for the future, although some difficulties were encountered such as for example the separation of certain copper plates forming the junction, or a powdering phenomenon thus creating a short circuit and preventing current recovery. However, we generally succeeded in cutting the fiber material cleanly and regularly and thus obtained a large TVIP.

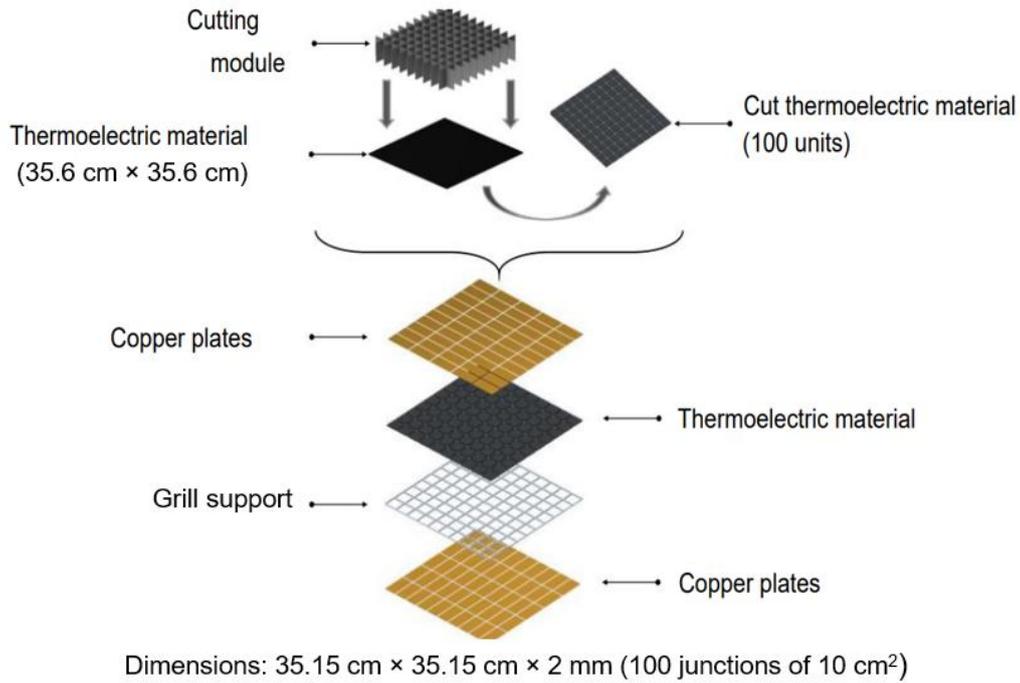

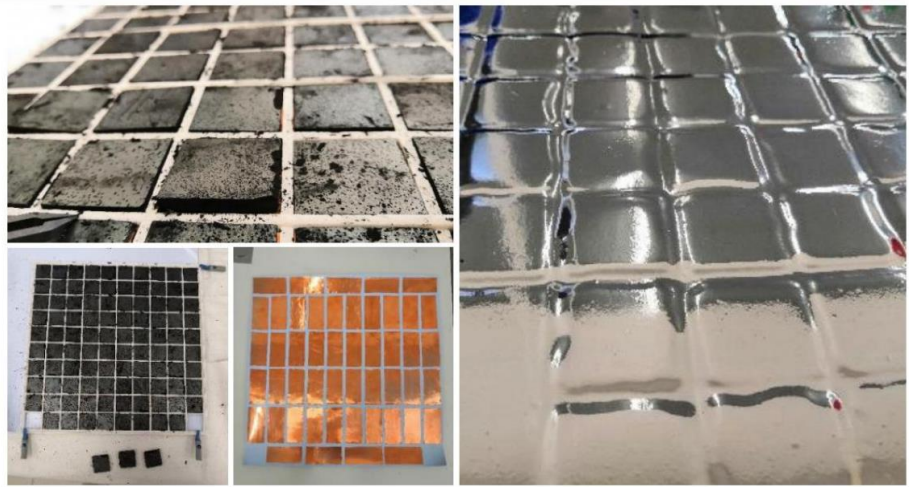

Figure 8: (A) CAD diagram of the assembly of a thermoelectric module with a surface area of 1000 cm², thickness 2 mm and composed of 100 junctions (B) Photos of the assembly of the thermoelectric module composed of 100 pads of approximately 1000 cm² surface area based on PANOX fiber carbogel material pyrolyzed at 850°

## 3. Discussion

Using a combination of pyrolysis and PANOX insertion, a 12-fold increase in electrical conductivity as well as in *ZT* were realized for RF xerogels whilst retaining their ultralow thermal conductivity (<50 mW/mK). Figure 9 summarizes the temporal evolution of the characteristics of the different families of thermoelectric materials with their figure of merit, their thermal conductivity, and their cost [2è, 29, 30, 32, 47-76]. While some other aerogels and polymers, p-PEDOT:PSS (no. 32), p-aerogel doped with carbon nanotubes (no. 33) and p-graphene aerogel (No. 34), offer higher *ZT*s, their higher costs and thermal conductivity limit their industrial scale application as thermal super insulating materials for surface applications, with such doping contents. Indeed, the cost price of such a material would be too high with a view to thermal insulation over a large area with an insulation thickness of 1 cm. Our material, RF-carbogel (no. 36), offers the best *ZT* for a thermal conductivity less than 50 mW.m$^{-1}$.K$^{-1}$, a density less than 120 kg .m$^{-3}$, with a cost less than 20 \$/kg. Evidently, this material is the best candidate if we want to move towards an application of thermal insulation combined with thermoelectric functionalization. This study allowed us to develop materials meeting such an application and having a lower cost than other aerogels and polymer-type materials encountered in the literature.

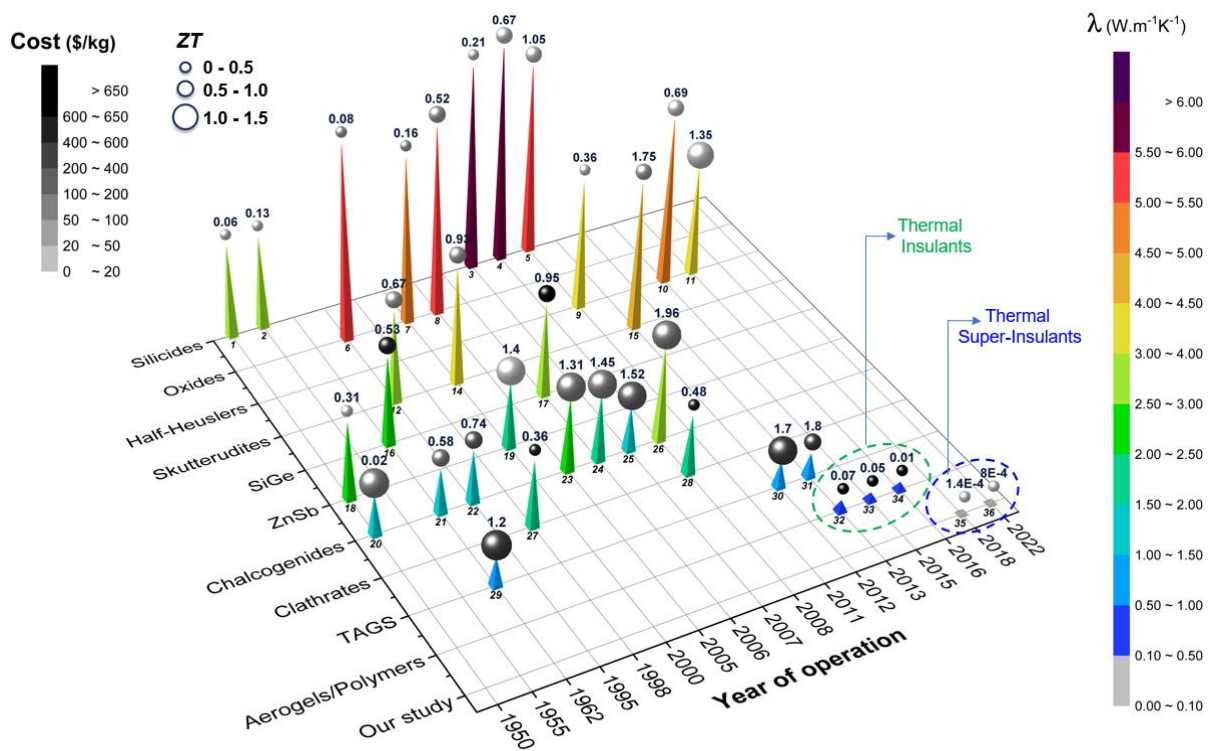

Figure 9: Temporal evolution of the characteristics of the different families of thermoelectric materials with their figure of merit, their thermal conductivity and their cost.

An assembly of the thermoelectric module, in the form of a vacuum insulated panel (VIP), was then carried out, using only a p-type leg with RF-carbogel, followed by the creation of the test bench. We demonstrated the use of this material in a proof-of-concept device application through an example of a failure detection system in the case of the battery of a hybrid vehicle. Importantly, such a system can simultaneously be used as a predictive sensor of battery failure due to thermal runaway because of its sensitivity to temperature. It is entirely conceivable to use this material in other thermal insulation applications under confined spaces. Successful repeatability proves the durability of the device for long-term use.

Lastly, the commercial prospect of such a device was explored by extrapolating the power output and voltage of the material in its current as well as optimized form. We considered the design of a thermoelectric module with junctions, first with exclusively p-p type junctions with the current material, then with p-p and n-p type junctions with materials having the thermoelectric properties that we foresee to achieve in the future. These properties can be achieved in various ways, either via the substitution of RF precursors by other types of electrically conductive organic materials (such as PEDOT: PSS, for example), or by modifying the synthesis parameters or even by using other types of conductive charges. Such optimization would then allow access to power density values of several Watts per square meter as well as output voltage values of several volts for small temperature differences. These values would therefore be very interesting for room temperature energy storage applications aimed at powering different remote wireless sensors or auxiliaries [42]. One application can be for certain difficult to access areas, e.g. as part of the insulation of an aircraft cabin (see figure S15 in the SI) where space constraints are high with significant constraints in terms of volume and weight. The little space available therefore has an impact on the technical specificities of the product (thin thickness, low thermal conductivity, etc.). This system can integrate into the required thermal insulation of the aircraft costing almost nothing in terms of weight and volume whilst providing complimentary applications like standalone sensors for the recovery of flight data (pressure, temperature, etc.) or thermal comfort through the Peltier effect where large temperature differences (-50°C exterior / +25°C interior) are at stake.

Other more general applications, e.g. in cold storage, pipelines, etc. can also be envisaged for both powering sensors and thermal management. Finally, a CAD study making it possible to demonstrate the feasibility of assembling such a large module was carried out along with the successful assembly from an experimental point of view. The key idea we want to convey is that low power density of carbogel based TE modules can be compensated by flexible large surface areas whilst their primary use as super-insulating material is further complimented by thermoelectric power generation.

The traditional approach has always been to start with a good TE material and then optimize commercial, environmental, and engineering factors for large-scale ambient heat recovery. Contrarily, here we present an alternative strategy of starting with a light thermal insulator, with insulation as its main function, and then optimizing its TE properties for complimentary heat recovery, resulting in environmental gain at almost no extra cost with no further optimization. Thus, our work opens a new avenue of using commercially available cheap carbogels with safe and earth-abundant elements, for sustainable and scalable ambient temperature waste heat recovery.

## 4. Methods

**4.1 Synthesis protocol**

The synthesis of our RF xerogel using acid catalysis and evaporative drying process was carried out according to the following steps:

- 20.38 g of resorcinol (R), (Fisher, 98%) are first dissolved in 200 ml of water in a cups of 500 ml.

- 5.0 g of pDADMAC (P) are added to the R + water mixture.

- After magnetic stirring, the pH is generally close to 6. It is then adjusted to 2 by slow addition of a catalyst solution (C) of HCl (10 M).

- Then 30.02 g of formaldehyde (F), (Aldrich, 37% by weight in water, stabilized by 10-15% of methanol) are added to the mixture.

- The solution is then mixed by magnetic stirring before being poured into a mold which is subsequently closed hermetically.

- Subsequently, the mold is placed in an oven at 90°C for 24 hours.

- This mold is then opened and the aqueous gel obtained is then dried by evaporation in an air oven at 90°C for another 24 hours.

- At the end of these steps, the RF xerogel is obtained in the form of a plate corresponding to a mold with a thickness of 1 cm and a side of 15 cm.

These steps are summarized through the summary diagram below (Figure 10):

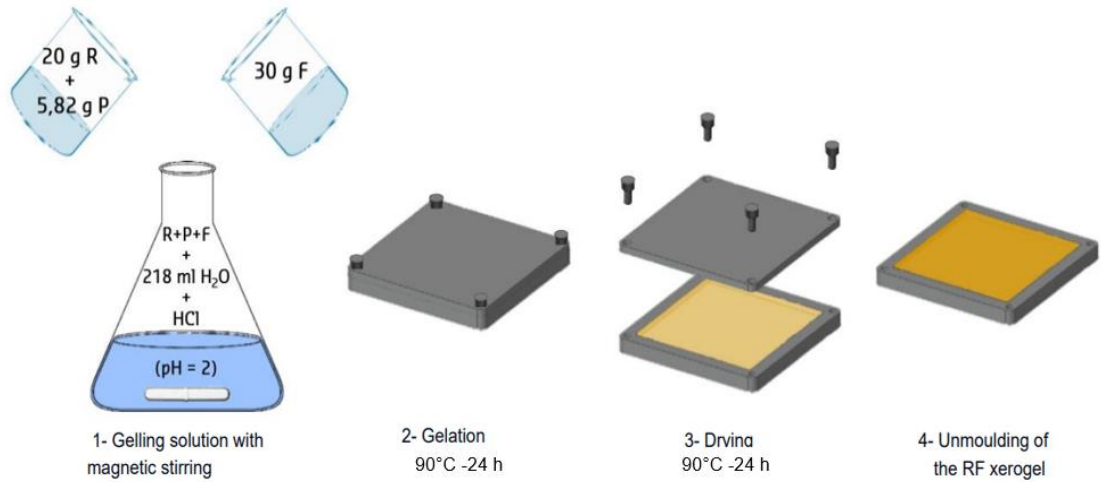

Figure 10: RF xerogel manufacturing process

The advantage of this synthesis process lies in the possibility of formulating RF xerogel plates with the desired dimensions and shapes, on a large scale, provided that the synthesis parameters detailed above are respected.

**4.2 Heat treatment by pyrolysis**

The pyrolysis step allows us to obtain a carbonaceous material via almost total elimination of heteroatoms such as N, O, H or S. It is carried out in a temperature range between 600°C and 1200°C under a flow of inert gas to prevent the combustion of the xerogel. At the end of this process, the carbonizate or carbon xerogel or carbogel obtained is made up of more or less ordered graphitic crystallites (polyaromatic carbons). During this work, two pyrolysis temperatures were investigated, 650°C and 850°C. After drying, the gels are pyrolyzed under a stream of nitrogen in a pyrolysis oven with a heating ramp of 1°C min$^{-1}$ in order to preserve the structure of the xerogel as much as possible. The chosen pyrolysis program also includes isotherm stages, namely a 1-hour stage at 140°C to ensure that all the water molecules absorbed on the surface are evaporated, then a second 30 min stage at 140°C, final temperature before cooling under nitrogen to room temperature. Note that beyond 600°C the $CH_2$ type bonds as well as the -$CH_2$-O-$CH_2$- bridges are eliminated [77]. The steps of one of our pyrolysis programs is illustrated in figure 11.

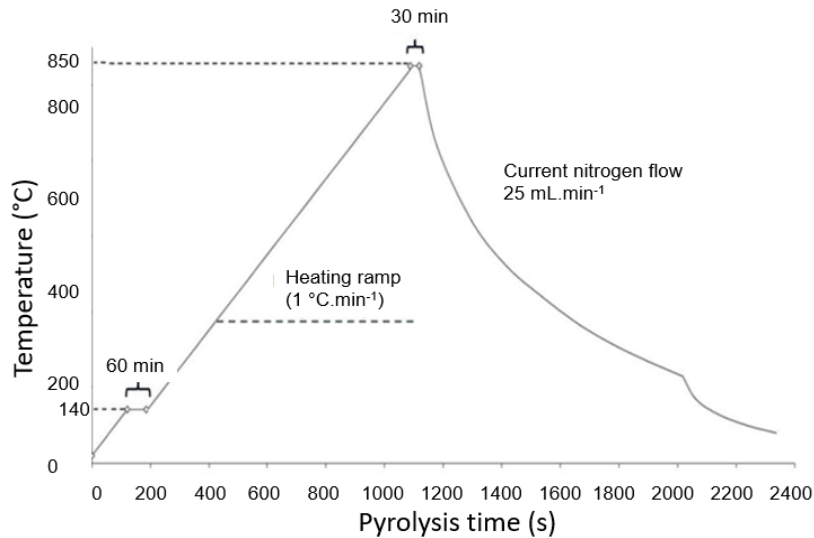

Figure 11: Profile of the pyrolysis program at 850°C used for the carbonization of the RF xerogel

### 4.3 Insertion of conductive fibers

As part of this study, we tried two types of conductive charges- basalt fiber (480 g.m$^{-2}$) and oxidized polyacrylonitrile fiber (PANOX 450 g.m$^{-2}$). They are characterized by their low density, good flexibility, thermal stability and good electrical conductivity. The basalt and PANOX fibers (see figure S14 in the SI) were purchased from the commercial sites of the Ferlam and SGL companies [78, 79]. Using an ultrasonic mixing process with the UP200St device from Hielsche, a homogeneous aqueous dispersion was obtained by slow addition of these conductive charges. This was followed by the acid catalysis synthesis (pH = 2) (see section 4.1), to obtain the doped RF carbogel. For each conductive filler, several doping rates were tested in order to reach the percolation threshold which describes the transition from an electrical insulating system to an electrically conductive system. Indeed, with a sufficient quantity of charges, charge carriers can move both by contact between particles as well as by tunneling [80].

### 4.4 Characterization

The thermal conductivity was measured with the HFM 436 flux meter from NETZSCH, the electrical conductivity was measured with the MTZ-35 impedance spectrometer from BioLogic and the Seebeck coefficient was measured by the ZEM-3 device from Ulvac Riko. A detailed method for thermoelectric characterization can be found in [12]. The density of the different materials was measured after drying via weighing on an electronic precision balance. A Gemini 300 scanning electron microscopy device from ZEISS was used to study the structure and topography of our materials at the nanometric scale with high resolution. This device was then coupled to Bruker's XFlash Detector 5030 to perform EDX elemental analysis to determine the chemical composition.

### 4.5 Manufacturing of the thermoelectric module

Developing the complex architecture of TEGs with carbogels is challenging due to its fragility. We addressed this challenge by developing the modules with a single type of material (p-type). The loss of yield caused by avoiding junctions with two types of materials (n-type and p-type), was compensated by our ability to design modules with a large size of 30 cm which otherwise is difficult for usual commercial materials due to their complexity of synthesis and greater rigidity. Another reason for this choice is to retain the primary super-insulating function of the material by obviating the need for thermal bridges, hence, avoiding the increase in thermal conductivity and the weakening of the material. To establish the electrical contact, two copper plates (supplied by GoodFellow) of 50 µm thickness were soldered with ultra-thin (<250 µm) flat copper electrical cable and covered with an adhesive that can withstand a temperature of up to 90°C. Then, in order to allow the module to be placed under vacuum while protecting the material and avoiding short circuits, we used a pocket of electrical insulating film (supplied by Rexor), transversely covering the entire material and leaving the passage for the electrical output cables. Lastly, the material was placed under vacuum in a waterproof envelop- thermoelectric vacuum insulation panel (TVIP)- to further reduce the thermal conductivity (< 15 $mW.m^{-1}.K^{-1}$) as well as contact resistance for allowing the participation of the material with entire surface roughness in the electricity generation. The welding was carried out by the vacuum machine at more than 100°C that melted the adhesive around the cable, making it possible to obtain a homogeneous and waterproof weld all around the copper electrical cable. The end of the two copper cables outside the casing is then connected to the initial electrical cables in order to facilitate the use of the module. The following figure represents the different elements constituting the TVIP (figure 12). A sealing test was conducted to confirm the vacuum retention of the TVIPs by comparing the evolution of the pressure in the envelope containing the thermoelectric module using standard electrical cables (leading to a rapid loss of vacuum) and that in the envelope with the thermoelectric module using electrical cables with a thickness of 250 µm. The results are presented in figure S3 and discussed in supplementary methods in section S2.2 of the SI. After fabricating the TVIP (see fig. S4), it is also necessary to create an operational test bench and delineate the formalism to evaluate the performance of Seebeck effect TE modules. This is discussed in section S1.3 and S1.4 in the SI.

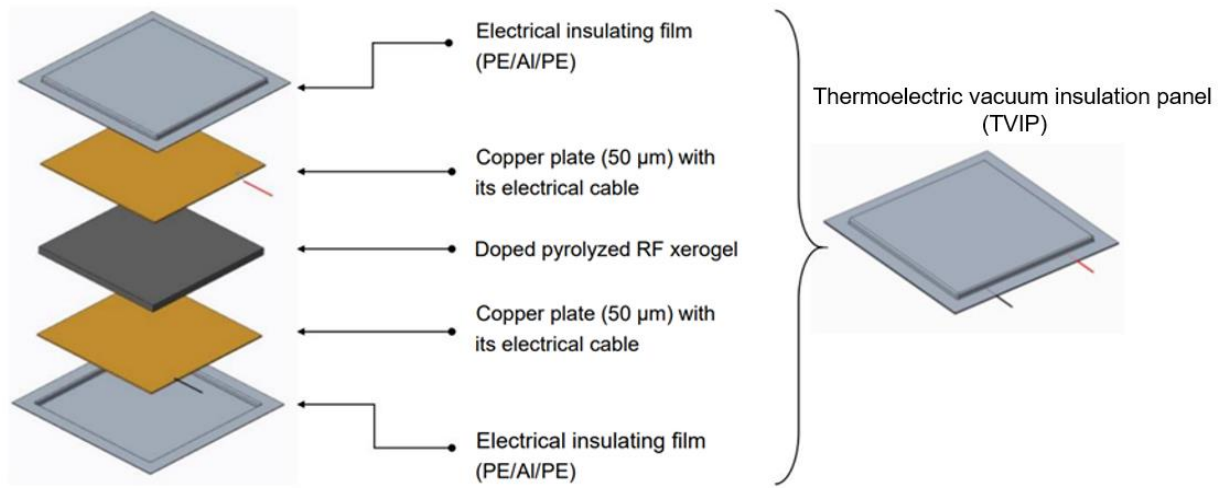

Figure 12: Diagram representing the different elements constituting the vacuum thermoelectric module

## Acknowledgments

Shoeb Athar and Jeremy Guazzagaloppa acknowledge the financial support from the Hutchinson SA (Châlette-sur-Loing, France).

## Competing Interests

A provisional patent based on this work has been filed by Jérémy GUAZZAGALOPPA and Cédric HUILLET. The other authors declare no Competing Financial or Non-Financial Interests

## Graphical Abstract

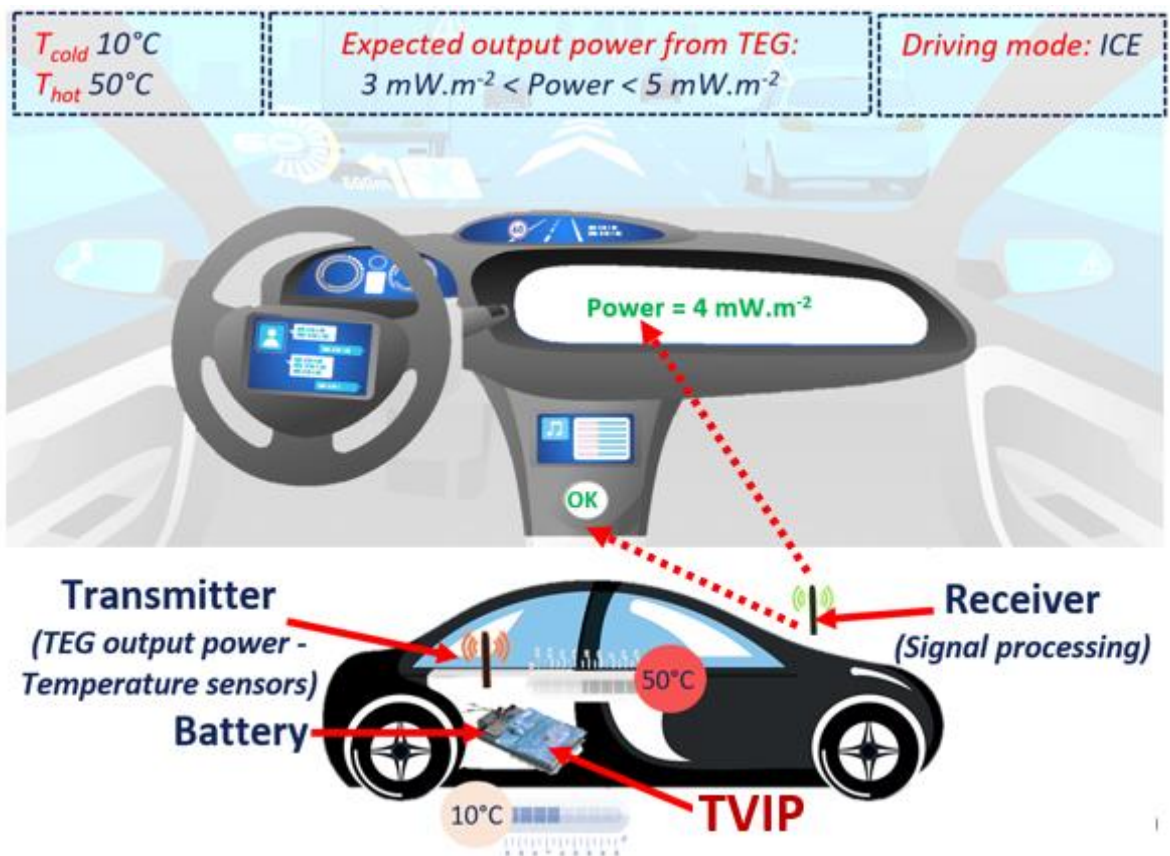